\documentclass[journal,letterpaper]{rmaa}

\usepackage{geometry}
\begin{document}

\title{Spectroscopic analysis of four post-AGB candidates} 

\author{R. E. Molina,\altaffilmark{1} S. Giridhar,\altaffilmark{2} 
C. B. Pereira,\altaffilmark{3} A. Arellano Ferro,\altaffilmark{4,5} and
S. Muneer\altaffilmark{6}
\medskip }

\altaffiltext{1}{Laboratorio de Investigaci\'on en F\1sica Aplicada y Computacional,
Universidad Nacional Experimental del T\'achira, CP 5001, Venezuela.}
\altaffiltext{2}{Indian Institute of Astrophysics, Bangalore 560034, India}
\altaffiltext{3}{Observatório Nacional/MCTI, Rua Gen. José Cristino 77, Sao Cristovao, Rio
de Janeiro, CEP 20921-400, Brazil}
\altaffiltext{4}{Instituto de Astronom\'{\i}a, Universidad Nacional Aut\'onoma de M\'exico,
M\'exico.}
\altaffiltext{5}{European Southern Observatory, Karl-Schwarzschild-Stra$\beta$e 2,
85748, Garching bei M\"{u}nchen, Germany}
\altaffiltext{6}{CREST Campus, Indian Institute of Astrophysics, Hosakote, India}

\fulladdresses{
\item R. E. Molina: Laboratorio de Investigaci\'on en F\1sica Aplicada y
Computacional, Universidad Nacional Experimental del T\'achira,  Venezuela,
(rmolina@unet.edu.ve). 
\item S. Giridhar: Indian Institute of Astrophysics, Bangalore 560034, India
(giridhar@iiap.res.in).
\item C. B. Pereira: Observatório Nacional/MCTI, Rua Gen. José Cristino 77, Sao Cristovao, Rio
de Janeiro, CEP 20921-400, Brazil (claudio@on.br).
\item A. Arellano Ferro: Instituto de Astronom\1a, UNAM, Apartado Postal 70-264, 
04510, M\'exico, D. F., M\'exico (armando@astro.unam.mx).
\item S. Muneer: CREST Campus, Indian Institute of Astrophysics, Hosakote, India
(muneers@iiap.res.in).}

\shortauthor{Molina et al.}
\shorttitle{PAGB candidates with selective depletion}

\resumen{Hemos efectuado un an\'alisis detallado de las abaundancias qu\1micas 
de cuatro objetos candidatos a post-AGB; IRAS 13110\,-\,6629, IRAS 17579\,-\,3121,
IRAS 18321\,-\,1401 y IRAS 18489\,-\,0629, usando espectros de alta resoluci\'on.
Hemos construido las Distribuciones de Energ\1a Espectral (SEDs) para cada objeto
usando datos fotom\'etricos existentes, combinados con los flujos infrarojos.
Para todos los objetos en la muestra, las SEDs exhiben dos picos en la distribuci\'on
de energ\1a, con el pico IR bien definido, mostrando con ello la presencia de
material circunestelar. Las abundancias de CNO muestran la producci\'on de N
por la v\1a del ciclo CN, mientras que el cociente [C/Fe] observado se\~nala una mezcla de 
carbono producida por la combusti\'on del He aunque el cociente C/O se mantiene
menor que 1. Se observa un efecto moderado de separaci\'on gas-polvo en IRAS
18489\,-\,0629 y en IRAS 17579\,-\,3121 mientras que solo se observa una gran dispersi\'on en
la remoci\'on
selectiva en IRAS 18321\,-\,1401 y en IRAS 13110\,-\,6629, lo que indica que otros 
procesos afectan los patrones de abundancias observadas.}

\abstract{We have done a detailed abundance analysis of four unexplored
 candidate post- Asymptotic Giant Branch(AGB) stars IRAS 13110\,-\,6629, IRAS 17579\,-\,3121, IRAS 18321\,-\,1401
 and IRAS 18489\,-\,0629 using high resolution spectra. We have
 constructed Spectral Energy Distributions (SED) for these objects using the existing
photometric
 data combined with infrared (IR) fluxes. For all sample stars, the SEDs exhibit 
 double peaked  energy distribution with well separated IR peaks showing the
 presence of dusty circumstellar material. The CNO abundances indicate the
 production of N via CN cycling, but observed [C/Fe] indicates the mixing
 of carbon produced by He burning by third dredge up although C/O ratio
 remains less that 1. A moderate DG effect is clearly seen for IRAS 18489\,-\,0629
and IRAS 17579\,-\,3121 while a large scatter observed in
 depletion plots for IRAS 18321\,-\,1401 and IRAS 13110\,-\,6629 indicate the
 presence of other processes affecting the observed abundance pattern.}


\maketitle

\section{INTRODUCTION}
\label{sec:introd}

Post-AGB (PAGB) stars, which represent the late stages of evolution of
low and intermediate  mass stars (0.8 -- 8 M$_{\odot}$), are very important
diagnostic tools to understand the evolutionary processes including AGB
 nucleosynthesis which is a major source of production of
elements such as C, N, F, Al, Na and s-process elements. These products are dredged up
 to the surface due to mixing episodes thereby modifying the observed abundances
 of these stars. The synthesized elements are ejected into the Interstellar Medium (ISM)
 through strong mass-loss at the end of AGB evolution hence these stars
 significantly affect the chemical evolution of galaxies.
 In the super wind phase (at the end of AGB evolution), most of the outer stellar
envelope is lost;  consequently PAGB stars
 are surrounded by circumstellar envelopes and hence their spectral energy
 distribution generally exhibits two peaks. However, exceptional
 PAGBs  without circumstellar envelopes have been detected.

The PAGB evolutionary model proposed by Iben and Renzini(1983)
 had been improvised by Groenewegen \& de Jong (1993) and Boothroyd \& Sackmann (1999)
 through better approximations for quantities like mass-loss and mixing length.
 More comprehensive models including detailed calculations of thermal pulses
 have been presented in Karakas, Lattanzio \& Pole (2002), Herwig (2004) etc.
 A very comprehensive review of AGB evolutionary models for a full range
 of masses and metallicities can be found in Herwig (2005).

 The canonical definition of PAGB stars as objects showing
 strong enhancement of carbon (C/O $>$ 1) and s-process elements is met by relatively
 smaller fraction of stars. Evolutionary models show that objects with a
 small range in masses (1.8 to 4.0 M$_{\odot}$) show carbon and s-process element
 enrichment
 while those with lower mass exhibit only the effect of CNO processing.
 More massive stars (M$>$  4.0 M$_{\odot}$) undergo Hot Bottom Burning (HBB) where
 carbon produced by He burning is converted to N, hence C/O remains $<$ 1 but
 N is enhanced and transient production of Li is predicted.

 The observational and theoretical developments made in the last two decades
 on PAGBs has been summarized in the excellent review by Van Winckel (2003) 
 where the diverse chemical compositions observed for the central stars are
 compared with the predictions from evolutionary models.
 More recent developments based on extended sample of
 PAGBs, made using the  extended wavelength coverage
 have augmented our understanding of PAGB classes
 including the morphology and chemistry of circumstellar shells as
 described in reviews by Garc\'ia-Lario (2006) and Giridhar (2011).

While the surface compositions of C-rich and O-rich PAGBs can
 be explained with AGB nucleosynthesis, mixing events and mass-loss;
 it has been noticed that chemical peculiarities not attributable to
  AGB nucleosynthesis are observed in
 a large fraction of PAGB stars including RV Tauri stars which
 show  abundance patterns reflecting a systematic removal of condensable elements.
 Well known examples of such objects are HR4049, HD~52961, BD +39$^{o}$4926, HD
44179; a good summary could be found in Van Winckel (2003).
 The depletion of elements show strong correlation with condensation temperature
 T$_{C}$ (defined as the temperature at which half of a
particular element in a gaseous environment condenses into grains) ;
 those elements with higher T$_{C}$ (easily condensable) are
 heavily depleted  while those with lower T$_{C}$ are largely unaffected.
 This removal of condensable elements onto the dust grain
 is commonly referred as dust-gas  winnowing or DG-effect.

 The full mechanism of DG effect is not yet fully understood, but
it  is generally believed to operate in a dusty  circumbinary disks surrounding
these objects as initially suggested by Waters, Trams \& Waelkens (1992).
 From the study of SED of a large sample of PAGB stars De Ruyter et al.
 (2006) reported the presence of dust at or near sublimation temperature
 very close to the star for most depleted objects irrespective
of their temperatures and ascribed it to the presence of gravitationally
bound disk. With the help of high spatial resolution interferometry in
mid IR, dusty disks around depleted objects Red Rectangle, HR 4049 and
HD 52961 have been resolved (Deroo et al. 2007).

 Sumangala Rao, Giridhar \& Lambert (2012) have compiled the
stellar parameters and  abundances  for PAGB stars and
 find that  s-process enhanced group contains
 a  very small number of binaries. However the depleted group
 contains larger fraction of binaries in accordance with the hypothesis
 of dusty disks surrounding binary post-AGB stars.
 Sumangala Rao \& Giridhar (2014) have presented a similar compilation for RV Tauri
stars.
 These authors also revisit the boundary conditions for discernible DG effect
 discussed in earlier papers by Giridhar  et al. (2000, 2005).
The values of minimum temperature and intrinsic metallicities are very similar for
 PAGB and RV Tauri stars.

The goal of this paper is to enlarge PAGB sample by a study
 of four candidate PAGB stars chosen from  their location in the
two color IRAS diagram. We have carried out
detailed atmospheric abundance analysis  using high resolution spectra
for IRAS 13110\,-\,6629,
 IRAS 17579\,-\,3121, IRAS 18321\,-\,1401, IRAS 18489\,-\,0629,
 previously identified as
post-AGB/PN objects (Preite-Mart\'inez 1988; Garc\'ia-Lario et al. 1997; Pereira \&
Miranda 2007). We have also studied their spectral energy distribution to
understand the properties of circumstellar material surrounding them.

The structure of the paper is as follows: $\S$ 2 presents the sample selection.
We describe our observations and data analysis procedures in $\S$ 3. The estimation
 of atmospheric parameters, abundances and the errors associated with these
 parameters are presented, in $\S$ 4.
 The derived abundances for  individual objects are reported. In $\S$ 5 these
results are discussed. In $\S$ 6  and $\S$ 7 we present summary and conclusions.

\section {The sample}
\label{sec:sample}

\begin{figure}
\begin{center}
  \includegraphics[width=\columnwidth]{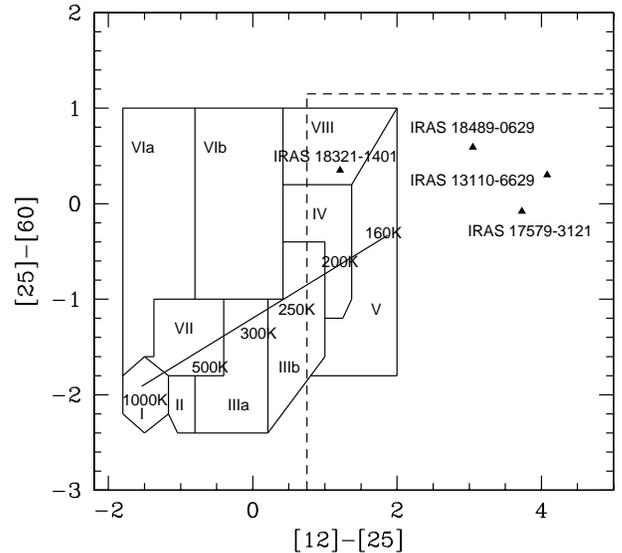}
  \caption{The location of our sample stars in IRAS two color diagram is shown. The
dotted-line
represents the region limit by Garc\'ia-Lario et al. (1997). The regions represented
with
Roman numerals have been defined by van der Veen \& Habing (1988).}
  \label{fig:figure1}
\end{center}
\end{figure}

Fig.~\ref{fig:figure1} shows the IRAS color-color diagram of van der Veen \&
Habing (1988)
with different zones signifying the emergence and evolution of the circumstellar shell
produced during AGB evolution. The objects in zone IV are undergoing super wind phase
or slightly beyond; objects with only cold dust are found in V while  VI b contains
objects with warm and cold shells. Zone VIII and dashed area  defined
by Garc\'ia-Lario et al. (1997) is a region where,  most
post-AGB and planetary nebulae (PN) are found.

We have selected sample stars whose IR colors places them in zone VIII (IRAS 18321\,-\,1401)
and the remaining three in the region where according to Garc\'ia-Lario et al. (1997)
PAGB and PPN are found.

IRAS 13110\,-\,6629 (GLMP 342) was included in a sample of candidate PN objects
by  Silva et al. (1993) attempting to detect OH (1612 MHz, 1665 MHz and 1667 MHz)
masers without success.
Its position in the IRAS color-color diagram $[12]-[25]$ =$+$4.08 and
$[25]-[60]$ = $+$0.30, indicates that this object falls into the region defined
by Garc\'ia-Lario et al. (1997) for PAGB stars and planetary nebulae. The presence
of circumstellar material was detected by Clarke et al. (2005). This object
has also been included in the Toru\'n catalog of Galactic PAGB  (Szczerba et al.
2007; Szczerba et al. 2011) as a likely PAGB object.

The star IRAS~17579\,-\,3121 (GLMP 686) was classified as candidate PN by
Preite-Mart\'inez (1988) and Ratag et al. (1990) due to its infrared fluxes.
Silva et al. (1993) reported the detection of a maser emission of OH at 1612 MHz
(with double peaks) where they found velocities for the blue peak of 21 km s$^{-1}$
and red peak 0 km s$^{-1}$ respectively. Su\'arez et al. (2006) classified this star
as a PAGB star from its optical counterpart based on a low resolution spectrum.
A more recent interferometric study using VLA by G\'omez et al. (2008) did not confirm
the association of IRAS 17579\,-\,3121 with the position of the radio continuum emission
and conclude that this IRAS source is not likely to be a PN.

Given its position in the IRAS two color diagram IRAS 18321\,-\,1401 (PM 1-243) 
was
classified as a possible PN by Preite-Mart\'inez et al. (1988). This classification
was questioned in a subsequent study by van de Steene \& Pottasch (1995) who employed
radio interferometric measurements to detect radio continuum emission at 6~cm
which is characteristics of PN; but no emission was detected for this object.
More recently, Pereira \& Miranda (2007) classified the star as a PAGB star by
comparing its spectrum with the spectrum of the known PAGB star GLMP 982.
These authors suggest a spectral type F for this object.

The star IRAS 18489\,-\,0629 (PM 1-261) has been classified as a probable
Planetary
Nebula by Preite-Mart\'inez (1988) following its location in the IRAS two color diagram.
The large value of [12]-[25] = $+$3.05 color \textbf{might indicate} the presence of
relatively warm dust resulting from a strong mass loss episode (Clarke et al. 2005).
Pereira \& Miranda (2007) from  their study of 16 PAGB candidates
using low resolution flux calibrated spectra have classified IRAS 18489\,-\,0629 as PAGB
since the spectrum closely resembles that of  GLMP 1058. By comparison
with the spectrum of the supergiant star HD 9973, taken from the library of Jacoby
et al. (1984),  Pereira \& Miranda (2007) suggested that the observed
continuum energy distribution of IRAS 18489\,-\,0629 was similar to that of
a reddened F5Iab star.

Basic information of the sample stars such as celestial and galactic coordinates,
brightness, IRAS fluxes and its quality (in parenthesis) is given in
Table~\ref{tab:table1}.
The apparent magnitudes $V$ were obtained only for IRAS 18489\,-\,0629,
IRAS 17579\,-\,3121 and IRAS 13110\,-\,6629 taken from different sources (Hog et al. 2000,
SIMBAD\footnote{http://simbad.u-strasbg.fr/simbad/sim-fid} database
and Perryman 1997).

\begin{figure}
\begin{center}
  \includegraphics[width=\columnwidth]{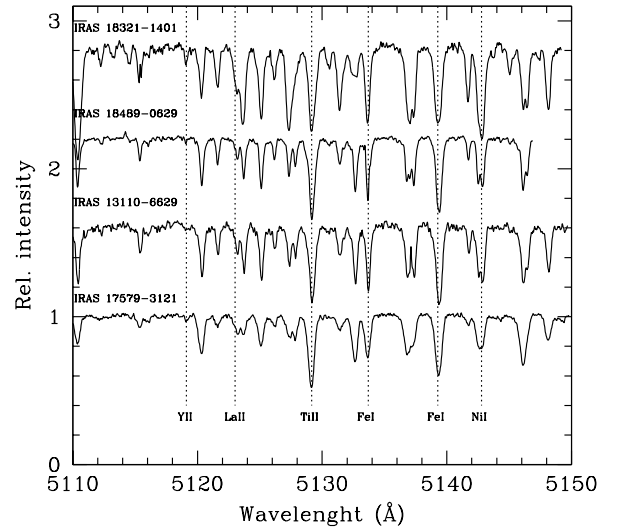}
  \caption{Representative spectra of the sample stars
 IRAS 13110\,-\,6629 and IRAS 17579\,-\,3121.
 IRAS 18321\,-\,1401, IRAS 18489\,-\,0629,
 The location of lines
 of certain important elements have been indicated by dashed lines.}
  \label{fig:figure2}
\end{center}
\end{figure}

\begin{table*}
 \centering
  \caption{Basic data of the program stars. The flux quality is given in parenthesis
coded from low to high as from 1 to 3 respectively.}
 \label{tab:table1}
\begin{tabular}{lccccccccc}
  \hline
  \hline
\multicolumn{1}{c}{No. IRAS}&
\multicolumn{1}{c}{$\alpha_{2000}$}&
\multicolumn{1}{c}{$\delta_{2000}$}&
\multicolumn{1}{c}{{\it V}}&
\multicolumn{1}{c}{{\it l}}&
\multicolumn{1}{c}{{\it b}}&
\multicolumn{1}{l}{F$_{12}$ $\mu$m}&
\multicolumn{1}{c}{F$_{25}$ $\mu$m}&
\multicolumn{1}{c}{F$_{60}$ $\mu$m}&
\multicolumn{1}{c}{F$_{100}$ $\mu$m}\\
\multicolumn{1}{c}{}&
\multicolumn{1}{c}{(h m s)}&
\multicolumn{1}{c}{(h m s)}&
\multicolumn{1}{c}{(mag)}&
\multicolumn{1}{c}{($^{0}$)}&
\multicolumn{1}{c}{($^{0}$)}&
\multicolumn{1}{c}{(Jy)}&
\multicolumn{1}{c}{(Jy)}&
\multicolumn{1}{c}{(Jy)}&
\multicolumn{1}{c}{(Jy)}\\
\hline
13110\,-\,6629&13 14 27.4&$-$66 45 35.0&10.74&305.20&$-$3.98&0.55(3)&23.51(3)&30.91(3)&12.48(3) \\
17579\,-\,3121&18 01 13.3&$-$31 21 56.5&11.40&359.61&$-$4.13&2.01(3)&62.39(3)&58.21(3)&14.71(1) \\
18321\,-\,1401&18 34 57.2&$-$13 58 49.0&$\ldots$&18.62&$-$2.75&0.36(1)&1.10(3)&1.52(3)&115.90(1) \\
18489\,-\,0629&18 51 39.1&$-$06 26 07.1&11.70&27.22&$-$2.97&0.27(1)&4.48(3)&7.49(3)&78.49(1) \\
\hline
\end{tabular}
\end{table*}

\subsection {Photometric Variability}
\label {sec:variability}

Light variability in PAGB stars is  of  common occurrence. RV Tauris exhibit
large amplitude light curve of distinct shape and periodicity  but other PAGBs are
known to exhibit small amplitude light variation as described by Kiss et
al. (2007).

PAGB stars of intermediate temperature are found near or above the upper
luminosity limit of the instability strip hence
semi-regular pulsations are commonly seen. Their periodicities are not always firmly
established and show irregular amplitude modulations and even cessation in their
variability. Classical examples of variable PAGB stars are 89 Her (HR 6685),
HD~161796 and LN Hya (HR 4912) with semi-periods of 63, 42-60 and 44-80 days (Arellano
Ferro 1985; Arellano Ferro 1981).

\begin{figure*}
\begin{center}
  \includegraphics[width=15.0cm,height=10.0cm]{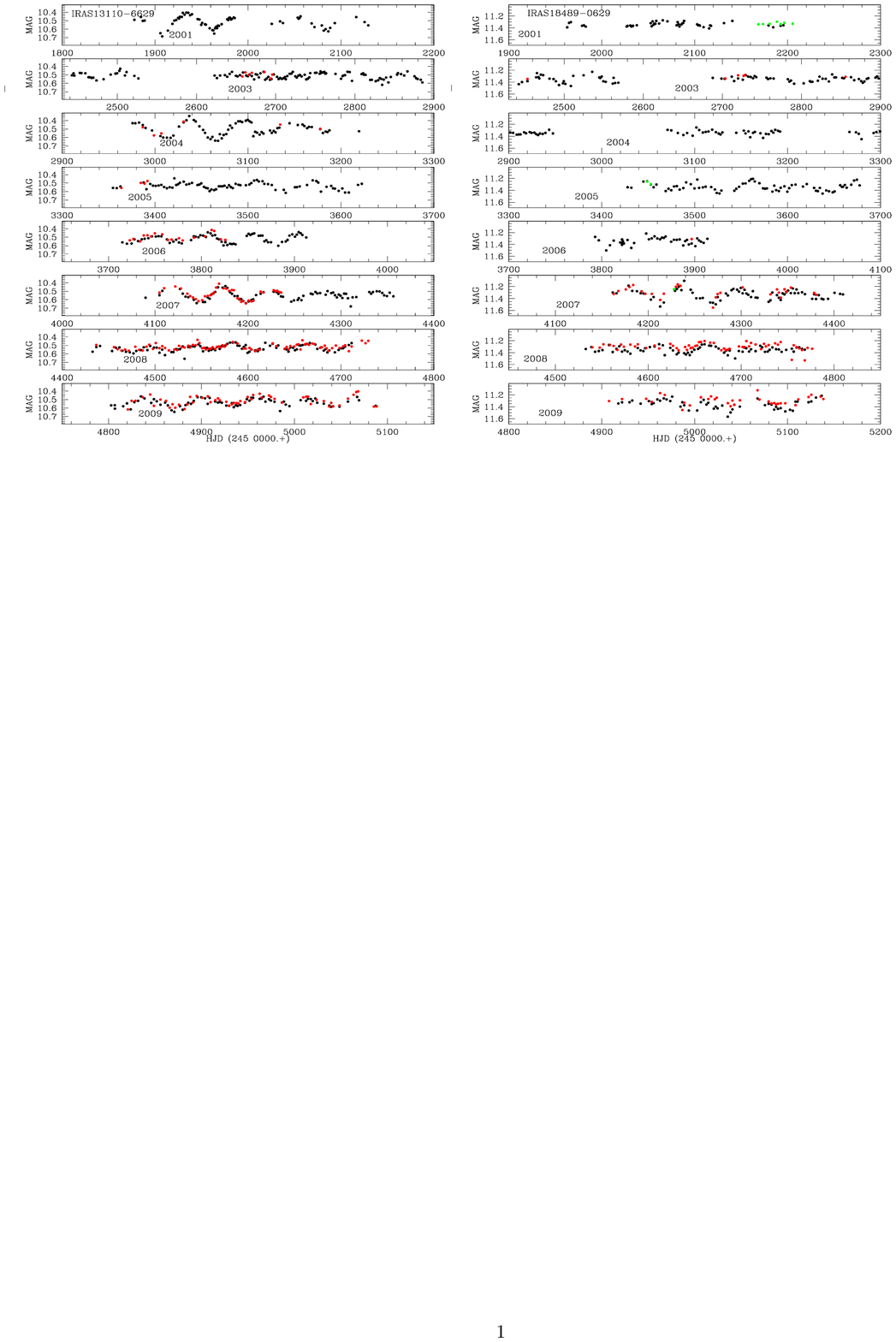}
  \caption{Light variations in the stars IRAS13110$-$6629 and IRAS18489$-$0629 as seen
from the V-band data from
the All Sky Automated Survey. Each independent data set is
depicted with a different Color.
Year numbers are placed at the beginning of the calendar year for a reference. See
$\S$ \ref{sec:variability}. for more details}
  \label{fig:lightcurves}
\end{center}
\end{figure*}

Defining periodicities  even using the radial velocity
variations for PAGB is complicated since their radial
velocities are often the combination of pulsations and differential
motions in the extended atmosphere (Kloshkova \& Panchuck 2012). On the other hand,
dense and continuous photometry of PAGB stars is rare.  In recent times the All Sky
Automated Survey (ASAS) (Pojmanski 2002) provided valuable photometric data for a
large number of objects.
We have examined the V-band data obtained between 2001 and 2009 for all four stars in
our sample.
IRAS13110\,-\,6629 and IRAS18489\,-\,0629 display clear signs of variability of
semi-regular nature, with amplitude modulations within 0.3 mag
(Fig. \ref{fig:lightcurves}). A frequency analysis of these data reveal
characteristic times or semi-period between 52 and 58 d for IRAS13110\,-\,6629 and
between
51.8 and 60.0 d for IRAS18489\,-\,0629. No secondary frequencies were identified. IRAS
17579\,-\,3121 and IRAS 18321\,-\,1401 do not show coherent variations but irregular
fluctuations that range between 0.2 and 1.0 mag.

\section{Observations and Reductions}
\label{sec:data}

The high-resolution spectra of the stars analyzed in this work
were obtained with FEROS, (Fiber-fed Extended Range Optical
Spectrograph) at the 2.2\,m ESO telescope at La Silla (Chile) on the
nights of August 27, 2007 (IRAS 18489\,-\,0629), May 14, 2009 (IRAS
17579\,-\,3121), July 29, 2009 (IRAS 18321\,-\,1401) and July 31, 2009
(IRAS 13110\,-\,6629). For each star two exposure times of 3\,600 secs
were obtained.  FEROS has a CCD with an array of 2048 $\times$ 4096
with each pixel size of 15$\mu$, manufactured by EEV.  The wavelength
range obtained with FEROS covers the spectral region between 3900 and
9200\AA\, and is distributed in 39 orders with a resolving power of
$\lambda$/$\Delta$$\lambda$\,=\,48\,000 or 2.2 pixels per
$\Delta$$\lambda$, which gives, at 5000\AA\, approximately 0.05\AA/pixel
(Kaufer et al. 1999). 

The spectra were reduced  following the standard procedure
including bias subtraction, flat-fielding, order
extraction and wavelength calibration with the MIDAS pipeline reduction package.

The cosmic ray hits causing narrow spikes were manually removed and
 all spectra were normalized to continuum. Two exposures were combined
to get higher S/N ratios.
The S/N ratio of the spectra was generally in the 120-150 range, however for
the spectrum of IRAS 13110\,-\,6629 it was about 70 because one of the exposures was
cut short due to cirrus. Since
most of the sample stars are of intermediate temperatures, line blending was
not severe. It was generally possible to measure line strengths with an
accuracy of 5 to 8 percent.

Fig.~\ref{fig:figure2} shows the representative spectra of the sample stars
arranged in descending temperature sequence.

\section{Atmospheric parameters}
\label{sec:param}

The spectral line strengths strongly depend on atmospheric parameters
such as effective temperature $T_{\rm eff}$, gravity log~$g$ and microturbulence
velocity $\xi_{t}$ in addition to the abundances; hence a good determination of
atmospheric parameters are mandatory to derive accurate abundances.
The estimation of the effective temperature and gravity can be made
from photometric and spectroscopic data.

\subsection{Photometric data}
\label{sec:photome}

Table~\ref{tab:table2} lists the magnitudes JHK for
IRAS 13110\,-\,6629 and IRAS 17579\,-\,3121 
 IRAS 18489\,-\,0629, taken from the 2MASS photometry (Cutri et al. 2003). 
The color excess $E(B-V)$ was estimated from the reddening map of Schlegel et al.
(1998).
For $E(B-V) > 0.15$, we adopted the correction suggested by Bonifacio, Caffau \&
Molaro (2000). Then the intrinsic colors $J_{0}$,
$(J-H)_{0}$ and $(H-K)_{0}$, were derived using the ratios
$\frac{E(J-H)}{E(B-V)} = 0.322$ and $\frac{E(H-K)}{E(B-V)} = 0.183$
of Fiorucci \& Munari (2003).

\begin{table*}
\tiny{
  \caption{2MASS photometry, intrinsic color and photometric atmospheric parameters
for the sample stars.}
  \label{tab:table2}
\centering
\begin{tabular}{lccccccccccc}
  \hline
  \hline
\multicolumn{1}{l}{No. IRAS}&
\multicolumn{1}{c}{$J\pm \sigma_J$}&
\multicolumn{1}{c}{$H \pm \sigma_H$}&
\multicolumn{1}{c}{$K \pm \sigma_{K}$}&
\multicolumn{1}{c}{$E(B-V)_{IS}$}&
\multicolumn{1}{c}{$(J-H)_{0}$}&
\multicolumn{1}{c}{$(H-K)_{0}$}&
\multicolumn{1}{c}{$J_{0}$}&
\multicolumn{1}{c}{$M_{J}$}&
\multicolumn{1}{c}{$T_{\rm eff,Tok}$}&
\multicolumn{1}{c}{$T_{\rm eff,Mol}$}&
\multicolumn{1}{c}{log~$g$,$_{Mol}$}\\
\multicolumn{1}{c}{}&
\multicolumn{1}{c}{(mag)}&
\multicolumn{1}{c}{(mag)}&
\multicolumn{1}{c}{(mag)}&
\multicolumn{1}{c}{(mag)}&
\multicolumn{1}{c}{(mag)}&
\multicolumn{1}{c}{(mag)}&
\multicolumn{1}{l}{(mag)}&
\multicolumn{1}{c}{(mag)}&
\multicolumn{1}{c}{(K)}&
\multicolumn{1}{c}{(K)}&
\multicolumn{1}{c}{}\\
\hline

13110\,-\,6629 & 8.06 & 7.60 & 7.40 & 0.50 & 0.30 & 0.15 & 7.62 &
$-$3.63$\pm$0.28 & 5864 & 6094$\pm$220 & 0.75$\pm$0.27 \\

17579\,-\,3121 & 8.396$\pm$0.017 & 7.890$\pm$0.046 & 7.585$\pm$0.021 & 0.60 & 0.312 & 0.195 & 7.863 &
$-$3.48$\pm$0.28 & 5722 & 6023$\pm$220 & 0.78$\pm$0.27 \\

18321\,-\,1401 & $\ldots$ & $\ldots$ & $\ldots$ & $\ldots$ & $\ldots$ & $\ldots$ & $\ldots$ &
$\ldots$ & $\ldots$ & $\ldots$ & $\ldots$ \\

18489\,-\,0629 & 9.006$\pm$0.027 & 8.63$\pm$0.05 & 8.398$\pm$0.029 & 0.51 & 0.213 & 0.139 & 8.560 &
$-$3.90$\pm$0.28 & 6733 & 6576$\pm$220 & 0.79$\pm$0.27 \\
\hline
\end{tabular}
}
\end{table*}

The effective temperatures  for program stars were first 
 derived using the $(J-H)_o$
calibrations for supergiant stars by Tokunaga (2000).
These temperatures are given in Table~\ref{tab:table2} as $T_{\rm eff,Tok}$. 

An independent estimates of temperatures, gravity and absolute magnitudes
were made  from intrinsic colors
$(J-H)_o$ and $(H-K_{s})_o$ calibrated by Molina (2012) for post-AGB and RV Tauri
stars. They are presented as
  $T_{\rm eff_{Mol}}$, log~$g_{Mol}$ and  $M_{J}$ in  Table~\ref{tab:table2}.
 As shown in the figure 6 of Molina (2012), the calibration made using F-G
 supergiants gives systematically lower temperatures.

These photometric estimates of atmospheric parameters serve as
starting values which are further refined by a detailed spectral analysis.

\subsection{Spectroscopic data}
\label{sec:spectro}

We have used new grids of ATLAS9 model atmospheres available at the database of 
Kurucz (see Castelli and Kurucz 2003).
We have used 2010 version of MOOG developed by C. Sneden (1973) in both line 
and spectrum synthesis mode. The assumptions made are local thermodynamic
equilibrium (LTE), plane parallel atmosphere and hydrostatic equilibrium.

The atmospheric parameters $T_{\rm eff}$, log~$g$ and the microturbulence velocity
$\xi_{t}$ are obtained from the lines of well represented elements such as Fe, Ti and
Cr.

The temperature was determined by requiring that there was no dependence of derived
abundances on lower excitation using Fe~I lines. Finally gravity was estimated from
the ionization equilibrium of Fe~I/Fe~II, i.e. $\log$ n(Fe II)$=$$\log$ n(Fe I).

Attempts have been made to get a unique solution for temperature
and gravity by studying the ionization equilibrium of several elements;
Mg~I \& Mg~II, Si~I \& Si~II, Ti~I \& Ti~II and Cr~I \& Cr~II.
The Balmer line profiles were affected by emission filling in hence
we have  explored the loci of Paschen lines in the temperature-gravity plane as
it will be illustrated later in this paper (see Fig. 5).

\begin{figure}
\begin{center}
  \includegraphics[width=8.cm,height=8cm]{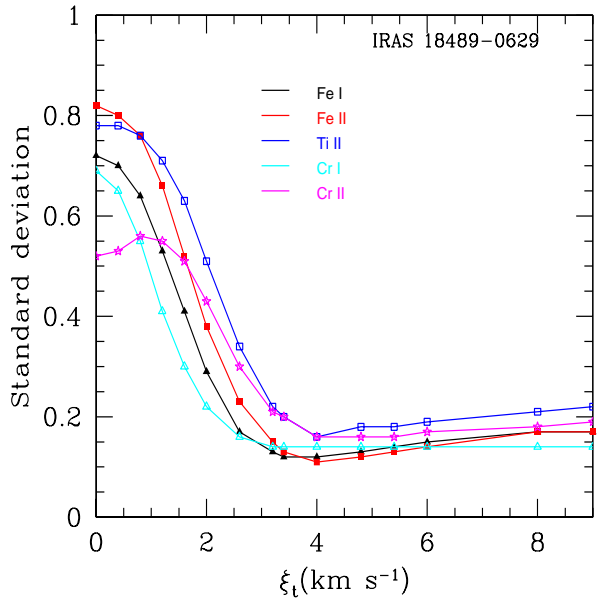}
  \caption{Determination of microturbulence velocity for the star IRAS
18489\,-\,0629 using the plot of standard deviation as a function of
microturbulence velocity $\xi_{t}$ for several species  in the 0 to 9 kms$^{-1}$ range.}
  \label{fig:turbulence}
\end{center}
\end{figure}

The microturbulence $\xi_{t}$ was estimated by requiring that the derived abundances
are independent of the line strengths for a given specie. Generally, lines of Fe~II
are preferred as it is known that they are not seriously affected
by departure from LTE (Schiller \& Przybilla 2008). For objects with
very few measurable Fe~II lines, Fe~I lines were employed.
In addition we have also employed the method described in
Sahin \&  Lambert (2009) for estimating the microturbulence velocity
based upon the measurement of
standard deviation as a function of microturbulence velocity $\xi_{t}$
for Fe I, Fe II, Ti~II, Cr~I and Cr~II lines in the 0 to 9 km~s$^{-1}$ range.   
Fig. \ref{fig:turbulence} illustrates the method using
the star IRAS 18489\,-\,0629 as an example. The minimum standard deviation is reached for
$\xi_{t} \sim 3.9$\,km~s$^{-1}$.

The measurement error in the equivalent widths (about 5 to 8 \%),
 corresponds to uncertainty in
microturbulence velocity   of $\pm$ 0.2 km s$^{-1}$  and in
temperature and gravity of
$\pm$ 200\, K and $\pm$ 0.20 respectively. The adopted values of $T_{\rm eff}$,
log~$g$ and $\xi_{t}$ for our sample are listed in Table \ref{tab:adopted}.

 \begin{table*}
 \centering
  \caption{Adopted atmospheric parameters for the program stars}
\label{tab:adopted}
\begin{tabular}{lcccrrr}
  \hline
  \hline
\multicolumn{1}{c}{No. IRAS}&
\multicolumn{1}{c}{$T_{\rm eff}$}&
\multicolumn{1}{c}{log~$g$}&
\multicolumn{1}{c}{$\xi_{t}$}&
\multicolumn{1}{c}{$V_{r}(hel)$}&
\multicolumn{1}{c}{$V_{r}(LSR)$}&
\multicolumn{1}{c}{Date}\\
\multicolumn{1}{c}{}&
\multicolumn{1}{c}{(K)}&
\multicolumn{1}{c}{}&
\multicolumn{1}{c}{(km s$^{-1}$)}&
\multicolumn{1}{c}{(km s$^{-1}$)}&
\multicolumn{1}{c}{(km s$^{-1}$)}&
\multicolumn{1}{c}{}\\
\hline
13110$-$6629 & 6500 & 0.50 & 5.0 & $+$11.5$\pm$0.5   & $+$8.1   & July 31, 2009\\
17579$-$3121 & 6500 & 0.00 & 4.7 & $+$24.7$\pm$0.6  & $+$34.3  & May 14, 2009 \\
18321$-$1401 & 5500 & 0.20 & 4.0 & $+$32.7$\pm$0.6  & $+$46.9  & July 29, 2009 \\
18489$-$0629 & 6500 & 0.50 & 3.9 & $+$159.6$\pm$0.5 & $+$175.3 & August 27, 2007\\
\hline
\end{tabular}
\end{table*}

\subsection {Uncertainties in the elemental abundances}
\label{sec:uncert}

The derived abundances are affected by errors in equivalent width
measurements (which in turn depends upon the spectral resolution,
S/N ratio, continuum placement and also on the spectral type of the star), errors
in atomic data such as oscillator strengths (log {\rm gf}) and errors
in deriving the atmospheric parameters. The equivalent
width measurement errors are random errors
while errors caused by uncertainty in log {\rm gf} and those in
estimating atmospheric parameters are systematic errors.
The effect of random errors in equivalent widths for a single star is well 
represented by $\sigma_{1}$, the standard
deviation from the mean abundance based on the whole set
of lines. These errors can be reduced by measuring a large number of lines.

The error in log gf values  vary from element to element. For example,
experimental values
for Fe~I and Fe~II of high accuracy, better than  5 \%, are
available for a large fraction of lines. For other Fe-peak
elements, errors in their ${\rm gf}$ values may range between 10 to 25\%.
For neutron-capture elements the accuracy of recent estimates are in 10\% to 25\%
range. An extensive list of  ${\rm gf}$ values for all important elements can be
found in  Sumangala Rao, Giridhar  \& Lambert (2012).
The effect of systematic errors in abundances (in  dex)
caused by errors in estimating the atmospheric
parameters are illustrated in Table~\ref{tab:table4}  for two representative stars
covering the temperature and gravity range of the sample stars.
We present change in abundances caused by varying atmospheric parameters
by 200\,K, 0.20 and 0.2 km s$^{-1}$ with respect to the chosen model for each star.

Following the standard procedure, the total systematic error $\sigma_{2}$ is
estimated by taking the square root of the sum of the squares
of each individual errors associated with uncertainties in temperature,
gravity and microturbulence.

 The total error $\sigma_{tot}$ for each element is quadratic sum
of  $\sigma_{1}$ and  $\sigma_{2}$. The error bars in abundance plot corresponds 
to this total error.

\begin{table*}
\caption{The sensitivity of abundances to the changes in the model atmospheric
        parameters for two values of temperature of our sample stars.}
\label{tab:table4}
\begin{center}
\begin{tabular}{lrrrrcrrrr}
\noalign{\smallskip}
\hline
\noalign{\smallskip}
\noalign{\smallskip}
   & \multicolumn{4}{c}{IRAS 18489\,-\,0629} & & \multicolumn{4}{c}{IRAS 18321\,-\,1401}\\
\cline{2-5}\cline{7-10}
   & \multicolumn{4}{c}{(6500\,K)} & & \multicolumn{4}{c}{(5500\,K)}\\
\multicolumn{1}{l}{Species}&
\multicolumn{1}{c}{$\Delta$ T$_{\rm eff}$}&
\multicolumn{1}{c}{$\Delta$ $\log$~$g$}&
\multicolumn{1}{r}{$\Delta$ $\xi_{t}$}&
\multicolumn{1}{c}{$\sigma_{tot}$}&
\multicolumn{1}{l}{}&
\multicolumn{1}{c}{$\Delta$ T$_{\rm eff}$}&
\multicolumn{1}{c}{$\Delta$ $\log$~$g$}&
\multicolumn{1}{r}{$\Delta$ $\xi_{t}$}&
\multicolumn{1}{c}{$\sigma_{tot}$}\\
\multicolumn{1}{l}{}&
\multicolumn{1}{c}{$+$200~K}&
\multicolumn{1}{c}{$+$0.20}&
\multicolumn{1}{c}{$+$0.20}&
\multicolumn{1}{c}{}&
\multicolumn{1}{l}{}&
\multicolumn{1}{c}{$+$200~K}&
\multicolumn{1}{c}{$+$0.20}&
\multicolumn{1}{c}{$+$0.20}\\
           \noalign{\smallskip}
            \hline
            \noalign{\smallskip}
C  I & $-0.07$ & $-0.01$ & $+0.02$ & $0.07$ & & $-0.06$ & $-0.12$ & $-0.08$ & $0.15$\\
N  I & $-0.09$ & $-0.06$ & $+0.02$ & $0.11$ & & $+0.07$ & $-0.04$ & $ 0.00$ & $0.08$\\
O  I & $+0.02$ & $-0.04$ & $+0.02$ & $0.05$ & & $-0.02$ & $-0.03$ & $+0.01$ & $0.03$\\
Na I & $-0.17$ & $+0.05$ & $+0.01$ & $0.18$ & & $-0.05$ & $+0.01$ & $+0.01$ & $0.05$\\
Mg I & $-0.22$ & $+0.06$ & $+0.01$ & $0.23$ & & $-0.06$ & $ 0.00$ & $+0.02$ & $0.06$\\
Al I &         &         &         &        & & $-0.05$ & $+0.01$ & $+0.01$ & $0.05$\\
Si I & $-0.17$ & $+0.05$ & $+0.01$ & $0.18$ & & $-0.06$ & $+0.01$ & $+0.01$ & $0.06$\\
Si II& $+0.08$ & $-0.05$ & $+0.02$ & $0.10$ & &         &         &         &       \\
S  I & $-0.13$ & $+0.01$ & $+0.02$ & $0.13$ & & $+0.01$ & $-0.02$ & $+0.01$ & $0.02$\\
Ca I & $-0.24$ & $+0.05$ & $+0.02$ & $0.25$ & & $-0.08$ & $+0.01$ & $+0.03$ & $0.08$\\
Sc II& $-0.15$ & $-0.05$ & $+0.02$ & $0.16$ & & $-0.04$ & $-0.04$ & $+0.03$ & $0.06$\\
Ti I & $-0.27$ & $+0.04$ & $+0.01$ & $0.27$ & & $-0.12$ & $+0.01$ & $+0.01$ & $0.12$\\
Ti II& $-0.14$ & $-0.06$ & $+0.04$ & $0.16$ & & $-0.04$ & $-0.03$ & $+0.03$ & $0.05$\\
V  I &         &         &         &        & & $-0.12$ & $+0.01$ & $+0.01$ & $0.12$\\
V II & $-0.13$ & $-0.06$ & $+0.04$ & $0.15$ & & $-0.03$ & $+0.03$ & $+0.03$ & $0.05$\\
Cr I & $-0.25$ & $+0.04$ & $+0.02$ & $0.25$ & & $-0.09$ & $+0.02$ & $+0.03$ & $0.09$\\
Cr II& $-0.07$ & $-0.07$ & $+0.05$ & $0.11$ & & $-0.01$ & $-0.04$ & $+0.02$ & $0.04$\\
Mn  I& $-0.25$ & $+0.04$ & $+0.06$ & $0.26$ & & $-0.09$ & $-0.04$ & $+0.01$ & $0.09$\\
Fe  I& $-0.21$ & $+0.05$ & $+0.03$ & $0.22$ & & $-0.09$ & $+0.01$ & $+0.01$ & $0.09$\\
Fe II& $-0.06$ & $-0.05$ & $+0.05$ & $0.09$ & & $-0.05$ & $-0.04$ & $+0.04$ & $0.07$\\
Co  I& $-0.30$ & $+0.04$ & $ 0.00$ & $0.30$ & & $-0.10$ & $+0.01$ & $+0.01$ & $0.10$\\
Ni I & $-0.22$ & $+0.04$ & $+0.01$ & $0.22$ & & $-0.08$ & $+0.03$ & $+0.01$ & $0.08$\\
Ni II& $-0.07$ & $-0.06$ & $+0.01$ & $0.09$ & &         &         &         &       \\
Cu I & $-0.27$ & $+0.04$ & $+0.01$ & $0.27$ & & $-0.12$ & $+0.01$ & $+0.02$ & $0.12$\\
Zn I & $-0.22$ & $+0.05$ & $+0.02$ & $0.23$ & & $-0.08$ & $+0.01$ & $+0.05$ & $0.09$\\
Y  II& $-0.16$ & $-0.05$ & $+0.02$ & $0.17$ & & $-0.04$ & $-0.03$ & $+0.02$ & $0.05$\\
Zr II& $-0.16$ & $-0.06$ & $+0.01$ & $0.17$ & & $-0.04$ & $-0.03$ & $+0.01$ & $0.05$\\
Ba II& $-0.30$ & $+0.01$ & $+0.03$ & $0.30$ & & $-0.05$ & $-0.02$ & $+0.01$ & $0.05$\\
La II& $-0.24$ & $-0.03$ & $-0.01$ & $0.24$ & & $-0.05$ & $-0.03$ & $ 0.00$ & $0.05$\\
Ce II& $-0.22$ & $-0.04$ & $ 0.00$ & $0.22$ & & $-0.06$ & $-0.03$ & $+0.02$ & $0.07$\\
Nd II&         &         &         &        & & $-0.07$ & $-0.03$ & $+0.01$ & $0.07$\\
Eu II& $-0.21$ & $-0.03$ & $+0.01$ & $0.21$ & & $-0.06$ & $-0.03$ & $ 0.00$ & $0.06$\\
            \noalign{\smallskip}
            \hline
            \noalign{\smallskip}
\end{tabular}
\end{center}
\end{table*}

\section { Results for Individual stars}
\label{sec:abund}

 The observed elemental abundances of the stars are
 governed by the composition of the natal ISM, the nucleosynthesis
 and mixing process in the course of its evolution and in some
 cases external effects such as mass exchange. For certain PAGBs
 and RV Tauri stars the DG effect introduces abundance anomalies.

The atmospheric abundances of all represented elements are given in Table~
\ref{tab:abund} for the sample stars.

\begin{table*}
\scriptsize{
\caption{Elemental abundances for the sample stars.}
\label{tab:abund}
\centering
\begin{tabular}{lccrcccrccrccr}
\hline
   &   &\multicolumn{3}{c}{IRAS 13110\,-\,6629} &\multicolumn{3}{c}{IRAS
17579\,-\,3121}
       &\multicolumn{3}{c}{IRAS 18321\,-\,1401}&\multicolumn{3}{c}{IRAS
18489\,-\,0629}\\
\cline{3-14}\\
\multicolumn{1}{l}{Species}&
\multicolumn{1}{c}{$\log \epsilon_{\odot}$}&
\multicolumn{1}{c}{[X/H]}&
\multicolumn{1}{c}{N}&
\multicolumn{1}{r}{[X/Fe]}&
\multicolumn{1}{c}{[X/H]}&
\multicolumn{1}{c}{N}&
\multicolumn{1}{r}{[X/Fe]}&
\multicolumn{1}{c}{[X/H]}&
\multicolumn{1}{c}{N}&
\multicolumn{1}{r}{[X/Fe]}&
\multicolumn{1}{c}{[X/H]}&
\multicolumn{1}{c}{N}&
\multicolumn{1}{r}{[X/Fe]}\\
\hline
C I  &8.39&$-0.01\pm$0.13&18&$+0.27$ &$+0.07\pm$0.13&15&$+0.39$ &$-0.11\pm$0.10&
7&$+0.32$ &$-0.09\pm$0.10&18&$+0.27$\\
N I  &7.78&$+0.47\pm$0.14& 5&$+0.75$ &$+0.60\pm$0.07& 5&$+0.92$ &$-0.19\pm$0.16&
3&$+0.24$ &$+0.01\pm$0.07& 5&$+0.37$\\
O I  &8.66&$+0.04\pm$0.06& 3&$+0.32$ &$+0.29\pm$0.13& 3&$+0.61$ &$+0.10\pm$0.13&
3&$+0.53$ &$-0.18\pm$0.21& 4&$+0.18$\\
Na I &6.17&$-0.02\pm$0.05& 3&$+0.26$ &$+0.27\pm$0.08& 3&$+0.59$ &$-0.21\pm$0.09&
4&$+0.22$ &$-0.06\pm$0.10& 4&$+0.30$\\
Mg I &7.53&$+0.09\pm$0.15& 4&$+0.37$ &$+0.03\pm$0.09& 1&$+0.35$ &$+0.07\pm$0.10&
3&$+0.50$ &$-0.13\pm$0.14& 3&$+0.23$\\
Mg II&7.53&$+0.09\pm$0.20& 2&$+0.37$ &$+0.26\pm$0.09& 1&$+0.58$ &              &  &   
    &              &  &       \\
Al I &6.37&              &  &        &              &  &        &$-0.66\pm$0.06&
2&$-0.23$ &              &  &       \\
Si I &7.51&              &  &        &$+0.02\pm$0.09& 8&$+0.34$
&$-0.06\pm$0.10&18&$+0.37$ &$-0.03\pm$0.09&11&$+0.33$\\
Si II&7.51&$-0.06\pm$0.11&11&$+0.22$ &$+0.09\pm$0.04& 1&$+0.41$ &              &  &   
    &$-0.15\pm$0.04& 1&$+0.21$\\
S I  &7.14&$-0.04\pm$0.05&syn&$+0.24$&$+0.14\pm$0.10& 2&$+0.46$ &$-0.14\pm$0.13&
3&$+0.29$ &$+0.22\pm$0.12& 8&$+0.58$\\
Ca I &6.31&$-0.35\pm$0.13&16&$-0.07$ &$-0.52\pm$0.09& 9&$-0.20$ &$-0.62\pm$0.11&
8&$-0.19$ &$-0.52\pm$0.08&16&$-0.16$\\
Ca II&6.31&$-0.28\pm$0.04& 2&$ 0.00$ &$-0.44\pm$0.04& 1&$-0.12$ &              &  &   
    &              &  &       \\
Sc II&3.05&$-0.69\pm$0.17& 6&$-0.41$ &$-0.87\pm$0.12& 5&$-0.55$ &$-0.70\pm$0.15&
7&$-0.27$ &$-1.06\pm$0.12& 5&$-0.70$\\
Ti I &4.90&$-0.60\pm$0.06& 1&$-0.32$ &              &  &       
&$-0.67\pm$0.10&11&$-0.24$ &$-0.70\pm$0.16& 3&$-0.34$\\
Ti II&4.90&$-0.52\pm$0.10& 7&$-0.24$ &$-0.72\pm$0.10& 8&$-0.40$ &$-0.66\pm$0.13&
8&$-0.23$ &$-0.59\pm$0.13&12&$-0.23$\\
V I  &4.00&              &  &        &              &  &        &$-0.35\pm$0.21&
2&$+0.08$ &              &  &       \\
V II &4.00&$-0.69\pm$0.12& 2&$-0.41$ &$-0.87\pm$0.02& 1&$-0.55$ &$-0.38\pm$0.16&
2&$+0.05$ &$-0.40\pm$0.08& 3&$-0.04$\\
Cr I &5.64&$-0.37\pm$0.13& 8&$-0.09$ &$-0.54\pm$0.10& 4&$-0.22$
&$-0.54\pm$0.15&12&$-0.11$ &$-0.48\pm$0.17& 9&$-0.12$\\
Cr II&5.64&$-0.35\pm$0.16&12&$-0.07$ &$-0.38\pm$0.11& 6&$-0.06$
&$-0.58\pm$0.13&11&$-0.15$ &$-0.41\pm$0.18&12&$-0.05$\\
Mn  I&5.39&$-0.40\pm$0.17& 5&$-0.12$ &$-0.40\pm$0.14& 4&$-0.08$ &$-0.74\pm$0.12&
7&$-0.31$ &$-0.55\pm$0.12& 5&$-0.19$\\
Fe  I&7.45&$-0.28\pm$0.13&84&        &$-0.34\pm$0.14&73&        &$-0.45\pm$0.12&132&  
    &$-0.33\pm$0.12&55&       \\
Fe II&7.45&$-0.28\pm$0.11&19&        &$-0.29\pm$0.09&17&        &$-0.41\pm$0.10&17&   
    &$-0.39\pm$0.12& 9&       \\
Co I &4.92&$-0.45\pm$0.08& 1&$-0.17$ &              &  &        &$-0.31\pm$0.14&
4&$+0.12$ &$-0.33\pm$0.08& 1&$+0.03$\\
Ni I &6.23&$-0.26\pm$0.14&13&$+0.02$ &$-0.37\pm$0.09& 4&$-0.05$
&$-0.47\pm$0.10&34&$-0.04$ &$-0.42\pm$0.10&21&$-0.06$\\
Ni II&6.23&$-0.20\pm$0.12& 2&$+0.08$ &$-0.39\pm$0.04& 1&$-0.07$ &              &  &   
    &$-0.44\pm$0.04& 2&$-0.08$\\
Cu I &4.21&$-0.16\pm$0.04& 1&$+0.12$ &$-0.22\pm$0.10& 2&$+0.10$ &$-0.53\pm$0.04&
2&$-0.10$ &$-0.53\pm$0.04& 1&$-0.17$\\
Zn I &4.60&$-0.37\pm$0.07& 3&$-0.09$ &$-0.36\pm$0.08& 2&$-0.04$ &$-0.25\pm$0.14&
4&$+0.18$ &$-0.35\pm$0.10& 3&$+0.01$\\
Y II &2.21&$-0.96\pm$0.05& 3&$-0.68$ &$-1.33\pm$0.06& 4&$-1.01$ &$-1.28\pm$0.10&
5&$-0.85$ &$-1.41\pm$0.16& 4&$-1.05$\\
Zr II&2.59&$-0.67\pm$0.07& 2&$-0.39$ &$-1.42\pm$0.04& 1&$-1.10$ &$-0.93\pm$0.04&
1&$-0.50$ &$-1.28\pm$0.04& 2&$-0.92$\\
Ba II&2.17&$-0.40\pm$0.07& 1&$-0.12$ &$-0.85\pm$0.07& 1&$-0.53$ &$-1.55\pm$0.07&
1&$-1.12$ &$-0.84\pm$0.07& 1&$-0.48$\\
La II&1.13&$-1.06\pm$0.05& 1&$-0.78$ &              &  &        &$-1.05\pm$0.05&
1&$-0.62$ &$-0.53\pm$0.10& 3&$-0.17$\\
Ce II&1.58&              &  &        &$-1.48\pm$0.09& 1&$-1.16$ &$-0.99\pm$0.16&
4&$-0.56$ &$-1.02\pm$0.15& 2&$-0.66$\\
Nd II&1.45&              &  &        &              &  &        &$-1.02\pm$0.12&
8&$-0.59$ &              &  &       \\
Sm II&1.01&              &  &        &              &  &        &$-0.92\pm$0.09&
5&$-0.49$ &              &  &       \\
Eu II&0.52&$-0.22\pm$0.06& 2&$+0.06$ &$-1.21\pm$0.06& 1&$-0.89$ &              &  &   
    &$-0.45\pm$0.06& 1&$-0.09$\\
\hline
\end{tabular}
}
\end{table*}

\subsection {IRAS 13110\,-\,6629}
\label{sec:iras13110}

For IRAS 13110\,-\,6629 (GLMP 342) the atmospheric parameters $\xi_{t}$, 
T$_{\rm eff}$ were derived from
the study of a large number of Fe~I and Fe~II line; for the estimation of
 log~$g$ the ionization equilibrium between Fe~I \& Fe~II as well as that of
 Mg~I \& Mg~II, Ca~I \& Ca~II, Ti~I \& Ti~II, Cr~I \& Cr~II were used.
 The spectroscopically derived values T$_{\rm eff}$ 6500K, log~$g$ 0.5
 and $\xi_{t}$ of 5.0 kms$^{-1}$ as given in Table~\ref{tab:adopted}
 are used for abundance analysis.
 The temperatures derived from photometry are systematically lower; errors
 in reddening correction for object  surrounded by circumstellar material
 could be one of the possible reasons.
The heliocentric radial velocity $+$11.5$\pm$ 0.5 km s$^{-1}$ on HJD 2455043.479
 was derived using 167 lines.

This star also shows a marginal Fe deficiency with [Fe/H]
= $-$0.28.  Similar to IRAS 18489\,-\,0629, it exhibits a large
number of C, N lines but [N/Fe] is a little larger, $+$0.75.
Replenishment of C (used in CN processing) from triple $\alpha$
reactions is indicated by [C/Fe] of $+$0.27.
The C/O ratio is $\sim$ 0.5.
[$\alpha$/Fe] as estimated from Mg, Si and S is $+$0.27 which together
with small radial velocity does not make a very compelling case for thick
disk population. Fig.~\ref{fig:depletion} indicates a mild DG effect.

\subsection {IRAS 17579\,-\,3121}
\label{sec:iras17579}

For this star IRAS~17579\-\,3121 (GLMP 686) we estimated T$_{\rm eff}$=5722K from 
the Tokunaga (2000) calibrations and  T$_{\rm eff}$=6023 $\pm$ 220K, 
log~$g$=0.78 $\pm$ 0.27 and $M_J$=$-$3.48 $\pm$ 0.28 mag from the Molina (2012). 
From excitation equilibrium of Fe I and Paschen line fit we estimated a temperature 
of 6500K and the ionization equilibrium of Fe~I/Fe~II, Ti~I/Ti~II, Cr~I/Cr~II  
led to the estimate of log~$g$=0.0.

The photometric estimates of temperature are systematically smaller and log~$g$
 larger. Although light variations 
are not yet reported for this object but variability in parameters  observed
for PAGBs could in addition to the errors in
reddening estimate may possibly explain systematic differences in atmospheric
parameters.  The adopted parameters for IRAS~17579\,-\,3121 
are T$_{\rm eff}$=6500K, log~$g$=0.0 and
$\xi_{t}$=4.7 km s$^{-1}$ are used for deriving abundances.

\begin{figure*}
\begin{center}
  \includegraphics[width=15cm,height=15cm]{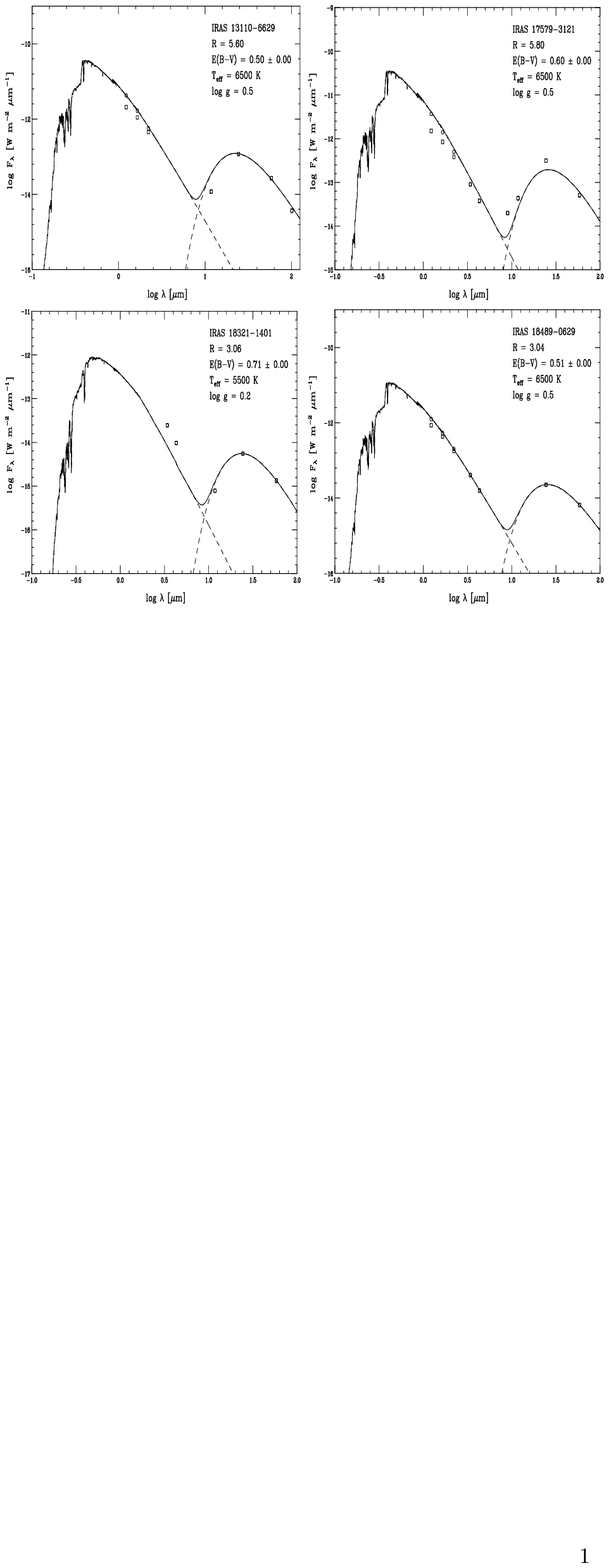}
  \caption{The SED for the program stars is generated
 using the existing photometric data and IR colors as described in the text.}
  \label{fig:SED}
\end{center}
\end{figure*}

\begin{figure}
\begin{center}
  \includegraphics[width=\columnwidth]{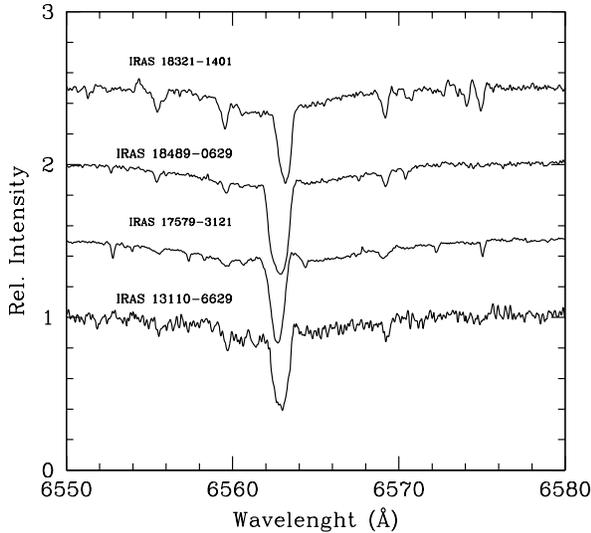}
  \caption{H$\alpha$ profiles in IRAS 18321\,-\,1401, IRAS 18489\,-\,0629, IRAS
17579\,-\,3121 and IRAS 13110\,-\,6629.}
  \label{fig:figure5}
\end{center}
\end{figure}

The heliocentric radial velocity measured for the spectrum taken on HJD 2454966.844
using  86 lines is $+$24.7 $\pm$ 0.6 km s$^{-1}$.

The photospheric abundances of IRAS 17579\,-\,3121
shows a moderate metal-deficiency ([Fe/H]= $-$0.32). Among light
elements, N shows significant enrichment [N/Fe] = $+$0.92 indicating
strong signatures of CN cycle; while [C/Fe] of $+$0.39 points to the
products of He burning being brought to the surface.
 We estimate [$\alpha$/Fe] of $+$0.4 similar to thick disk value.
 We again find the signature of DG effect as illustrated in
 Fig.~\ref{fig:depletion}.

\subsection {IRAS 18321\,-\,1401}
\label{sec:iras18321}

\begin{figure}
\begin{center}
  \includegraphics[width=8.cm,height=8cm]{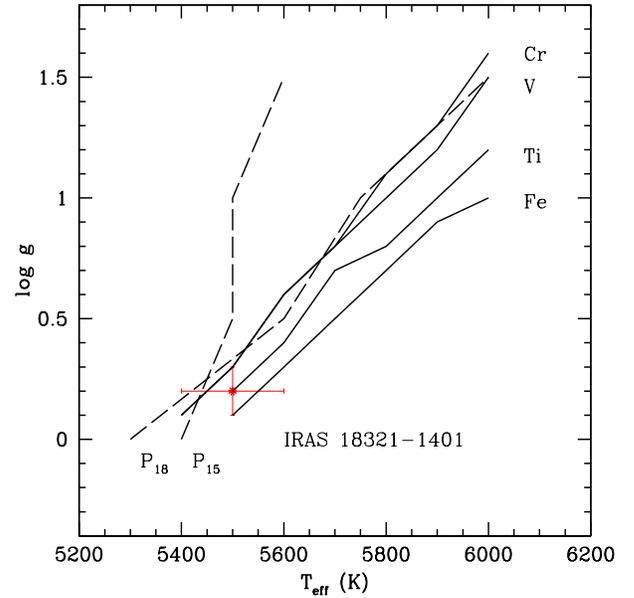}
  \caption{ The P15 (8545\,\AA) \& P18 (8438\,\AA) and the ionization equilibrium loci 
for Fe, Cr, V, Ti are plotted in the temperature gravity plane for IRAS18321\,-\,1401. 
The red dot indicates the adopted values of T$_{\rm eff}$ and log~$g$ for the calculation 
of abundances.}
  \label{fig:loci18321-1401}
\end{center}
\end{figure}

There is no $JHK$ photometry available for IRAS 18321\,-\,1401 to estimate the
initial temperature. H$\alpha$ and H$\beta$ being affected by emission were not used
as temperature indicators. 
However the matching of P15 (8545\,\AA) and P18 (8438\,\AA) profiles of Paschen
lines with 
those synthesized for a set of model atmospheres covering a good range in temperature 
and gravities and the combination with the ionization equilibrium requirement for Fe, 
Cr, Ti and V led to a temperature value of $\sim$5500 K and surface gravity of log~$g$=0.20.
Fig.~\ref{fig:loci18321-1401} shows the corresponding loci in the temperature-gravity
plane.

From our spectrum obtained on HJD 2455042.632, we measure a heliocentric radial
velocity of $+32.7 \pm 0.6$ km s$^{-1}$ using  141 clean lines.

The star is moderately metal-poor ([Fe/H]=$-$0.43). Like IRAS 13110\,-\,6629 and IRAS
18489\,-\,0629, IRAS 18321\,-\,1401 has a significant number of C~I lines. Very mild
enrichment of N and C is present. However this stars has smaller C/O
ratio of $\sim$0.3.
 We find [$\alpha$/Fe] of $+$0.38 similar to thick disk objects.
 This star also shows a moderate depletion
of condensable elements similar to that observed in IRAS 18489\,-\,0629 (see
Fig.~\ref{fig:depletion}). It should be noted that the scatter in the
plot is much larger indicating the influence of other parameter/process.

\subsection {IRAS 18489\,-\,0629}
\label{sec:iras18489}

For IRAS 18489\,-\,0629 (PM 1-261) the estimated values of the atmospheric 
parameters based upon a large number of lines covering a large
range equivalent widths, LEP and both stages of
ionization used for the abundance analysis are listed in Table \ref{tab:adopted}.
The temperature estimated by spectroscopy is in good agreement with that indicated
by the spectral type mentioned above as well as the photometric estimate
given in Table 2.

\begin{figure*}
\begin{center}
  \includegraphics[width=15cm,height=15cm]{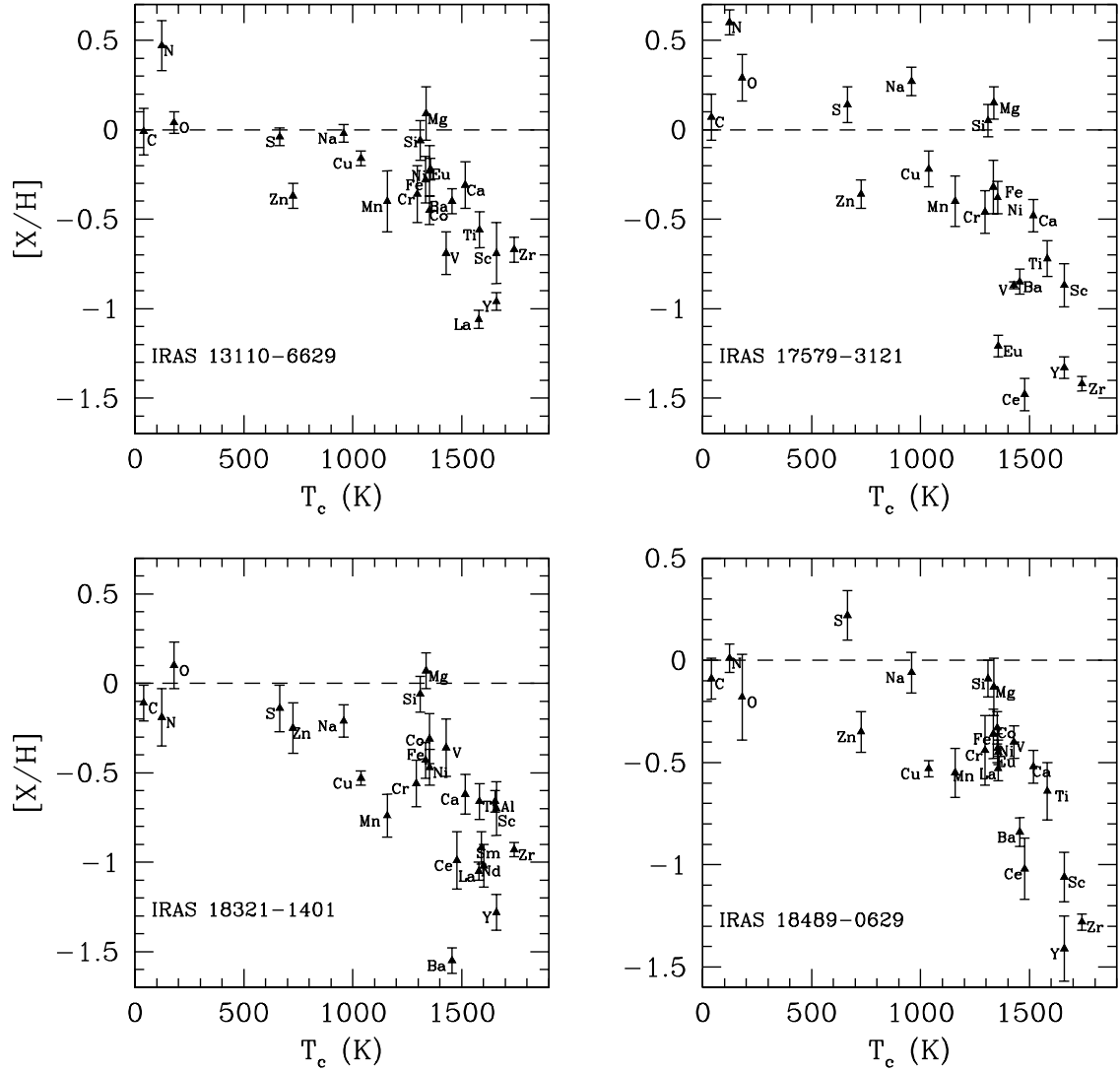}
  \caption{Elemental abundances [X/H] as a function of condensation temperature
T$_{c}$ (Lodders 2003) for the sample stars. }
  \label{fig:depletion}
\end{center}
\end{figure*}

From the spectrum taken on HJD 2454340.496 a heliocentric radial
velocity of $+159.6 \pm 0.5$ km s$^{-1}$ is measured using
223 clean unblended absorption lines that cover a wide spectral range.
 The present study is the first detailed abundance analysis made
for this star using  high resolution spectrum.
The spectrum contained a large number of C and N lines.
The star shows a mild iron deficiency ([Fe/H]= $-0.36\pm0.12$).
 From Table~\ref{tab:abund} it can be seen that not only 
 N shows small but significant enhancement indicating CN processing;
  C is replenished by the mixing of triple $\alpha$ products.
  We derive a  C/O ratio of  $\sim$ 0.8.
IRAS 18489$-$0629 shows a mild Na-enrichment, [Na/Fe] = $+$0.30, probably
caused by proton capture on $^{22}$Ne.
 Although our study covers five $\alpha$ elements Mg, Si, S, Ti and Ca,
 we employ only  Mg, Si and S to measure [$\alpha$/Fe] since
 Ti and Ca are susceptible to non-LTE and are also affected
 by DG effect. The derived [$\alpha$/Fe] of $+$0.38 together
 with large radial velocity makes it a likely thick disk object not withstanding
 a modest Fe deficiency.
 The Fe-peak elements V, Cr, Mn, Co, Ni vary in lockstep with Fe
 while s-process elements are significantly deficient.
The abundance pattern can be easily understood via  Fig.~\ref{fig:depletion}
 which clearly shows the dependence of [X/H] on T$_{C}$. IRAS 18489$-$0629
 belongs to the group of PAGBs showing systematic depletion
 of condensable elements.

\section {Discussion}
\label{sec:discussion}

In our study of unexplored PAGB candidates we find the signature of
 CN processing in all four sample stars although [N/Fe] ranges from
 $+$0.24 to $+$0.92. The measured [C/Fe] between +0.27 and +0.39
 clearly shows the mixing of He burning products via the third dredge up (TDU).
 The sample stars have SEDs displaying a double peak (see Fig.~\ref{fig:SED})
 and the observed [$\alpha$/Fe] in excess of $+$0.2 dex commonly seen
  in thick disk objects, which contain a large fraction of
  s-process enhanced PAGBs (see Table 9 of Sumangala Rao et al. 2012).
  However, none of the sample stars exhibit the expected enhancement of s-process
 elements. Instead, the observed [X/H] shows a dependence on T$_{C}$
 as demonstrated in (Fig.~\ref{fig:depletion}),
 although the extent of depletion varies over the sample.
 To quantify this effect, we could use the ratio of abundance of least
 affected elements like S and Zn (with low T$_{C}$), and the most affected ones such
as Sc, Al and s-process elements. 
The potential candidates for the representation of intrinsic
 (unaffected) metallicity are S and Zn. But S being an $\alpha$ element,
 shows relative enrichment for the thick disk and halo stars.
  Zn on the other hand, remains unchanged over a large range in metallicities.
Hence Zn is a better indicator of initial metallicity and has been used
  as reference to judge the relative enrichment of various elements.
  Among the elements most affected by the DG effect, Al and Zr are not
 well represented by many lines while Sc and Y measurements are based upon
 4-5 lines. Hence we chose to define DG index as [Zn/H] - ([Sc/H]+[Y/H])/2.

In our sample,
IRAS 18489\,-\,0629 with DG of 0.88 leads the group followed by IRAS 17579\,-\,3121
(0.74), IRAS 18321\,-\,1401 (0.67) and IRAS 13110\,-\,6629 (0.45) being the 
least depleted of the four. Another striking feature of the depletion pattern
 shown in Fig.~\ref{fig:depletion} is the large scatter found for
  IRAS 18321\,-\,1401 and IRAS 13110\,-\,6629. The intriguing aspect of the
  sample is that three objects showing a  range in
 DG index have  the same T$_{\rm eff}$ of 6500K and not withstanding its
 lower temperature, IRAS 18321-1401 does not have lowest DG index.
 All four sample stars have temperatures larger than lower
 temperature limit required to sustain the depletion pattern against
 the mixing caused in the extended convective envelopes of cooler stars, which
according to Sumangala Rao, Giridhar  \& Lambert (2012) is 4800K.

\subsection {The observed SEDs and H${\alpha}$ profile}
\label{sec:profileH}

 The SEDs for program stars were generated by comparing the archival 
 photometric data with the theoretical SED fluxes given in Kurucz (1991) 
 for models with atmospheric parameters given in Table 3
 and adopting the
  interstellar extinction model by Steeman \& The (1991).
 The ratio of total to selective extinction R= A$_V$/E(B-V) was estimated
 iteratively. We have used archival data from
  USNO, 2MASS, IRAS, AKARI, WISE and MSX6C available at NASA/IPAC
 Infrared Science Archive. We have omitted data with large errors
e.g. IRAS with quality flag =1, 2MASS quality flag = E/X/U/F, MSX6C quality flag =1.
The SEDs constructed for the program stars are shown in Fig.~\ref{fig:SED}.
   A Planckian fit to the observed fluxes between 1.0 to 200$\mu$m
 leads to dust temperatures of
 130K for IRAS 13110\,-\,6629, 110 K for IRAS 17579\,-\,3121,
 120K for IRAS18324\,-\,1401 and 110K for IRAS 18489\,-\,0629. 
 For IRAS 18321\,-\,1401 the scanty photometric data resulted in
  poor fit with the theoretically constructed SED.

Fig.~\ref{fig:figure5} shows the profiles of H$\alpha$ at 6562.8 \AA~ of the
program stars. All profiles have a broad shallow component and a deep
narrow component with significant asymmetry.
 All profiles show  indication of filling in by a rather central
emission. Incipient P-Cygni structure is also seen in all cases.
 The P-cygni profiles (or inverse P-Cygni profiles) are of common
 occurrence in PAGB stars and are ascribed to the presence of a shock
 propagation in the pulsating atmospheres of these stars.

\subsection {Overall abundance patterns}
\label{sec:Oabun}

We have observed a mild enrichment of [Na/Fe] $\sim$ $+$0.2 to
$+$0.3 in
IRAS 18489\,-\,0629, IRAS 13110\,-\,6629 and IRAS 18321\,-\,1401 and of $+$0.6 in IRAS
17579\,-\,3121. The enrichment of the latter object was measured from lines at
4979, 5682, 5688, 6160\,\AA, for which the non-LTE
corrections do not exceed $-$0.10 dex (Gehren et al. 2004; Lind et al. 2011).
This Na enhancement is probably due to a deep mixing that bring products of the
Ne-Na cycle to the surface.
 The [$\alpha$/Fe] measured using S, Si, Mg is similar to those found in
 thick disk objects although the observed metal deficiency is quite moderate.
 The observed [C/Fe] in the range $+$0.2 to $+$0.39 points to the mixing of He burning
products
 being brought to the surface but accompanying s-process enhancement either did not
 take place or was obliterated by the depletion.

 All four stars show depletion but a remarkable aspect of the abundance pattern is
 that only elements with T$_{C}$ greater than 1300K
 are significantly affected. 
It is not clear if the observed depletion pattern is caused by insufficient
 mixing of depleted material with photospheric gas or temperature and compositional
 structure of circumstellar material favored  condensation process only for
 high T$_{C}$ elements.

\begin{figure*}
\begin{center}
  \includegraphics[width=12.cm,height=10.cm]{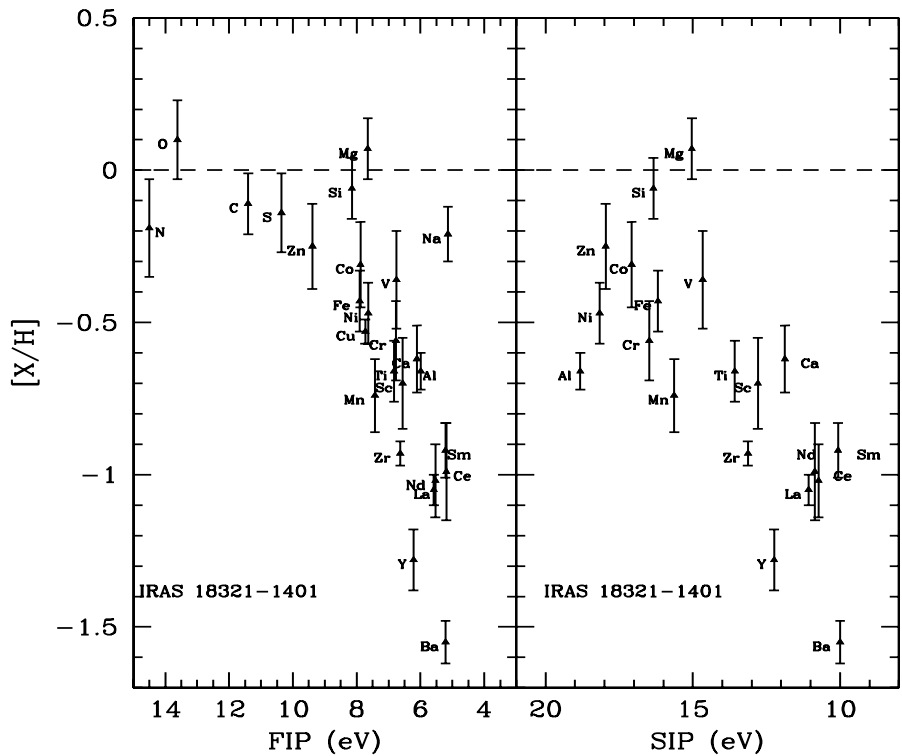}
  \caption{Abundances [X/H] versus FIP and SIP for IRAS18321\,-\,1401.}
  \label{fig:18321-1401}
\end{center}
\end{figure*}

To understand
 the large scatter present in the depletion diagram for IRAS18321\,-\,1401,
 we investigated the possible dependence of abundances on
 First Ionization Potentials (FIP) as suggested by Kameswara Rao \& Reddy (2005).
 A plot of observed abundances as a function of FIP
 left panel of  Fig.~\ref{fig:18321-1401} 
 clearly shows a reduced scatter. It appears that elements with FIP
 8eV or lower show systematic depletion.

 We also explored the possible dependence of
 observed abundance on the Second Ionization Potential (SIP)
 right column of   Fig.~\ref{fig:18321-1401} 
  and do notice a significant depletion of elements
 with SIP lower than 13.6eV. It may be recalled that
 Luck \& Bond (1989) had proposed that hydrogen
 Lyman continuum produced  from a shock in the atmosphere
 over ionized elements with a SIP lower than 13.6eV, the
 Lyman limit. The Ca, Sc, and s-process elements Y, Zr, Ba, La, Ce, Sm
 Eu have lower SIP and hence could be over ionized relative to
 LTE estimate hence the LTE abundance analysis would
 yield systematically low abundance of these elements.

 However, in the SIP plot the elements with similar SIP show
 different depletion. On the other hand,
 a smooth variation with FIP (with exception
of  Mg and Na) implies the systematic removal of species mostly present
  in first ionized state. The magnetically driven stellar wind
  proposed by Garc\'ia-Segura et al. (2005) for PAGB stars  has the potential to
  remove the ionized species,  but the presence of the stellar wind
  need to be established for this object. Although P-Cygni structure
  is mildly present in H$_{\alpha}$ the emission barely rises above continuum.
Na D indeed exhibits a complex structure. This star deserves a multi-wavelength
  monitoring to understand the observed complex abundance pattern.

\section {Conclusions}
\label {sec:conclusion}

In this paper, we have performed a detailed atmospheric abundance analysis from high
resolution spectra aimed to verify the evolutionary stage of a sample of four PAGB
candidates IRAS13110\,-\,6629, IRAS 17579\,-\,3121, IRAS 18321\,-\,1401 
and IRAS 18489\,-\,0629,
 selected from their infrared characteristics.
 None of them exhibit the s-process enhancement but clear indication of modest
 DG effect are observed for IRAS 18489\,-\,0629 and IRAS 17579\,-\,3121.
For IRAS 13110\,-\,6629
 and IRAS 18321\,-\,1401 the depletion diagram displays a large scatter.
 The scatter for IRAS 18321\,-\,1401 is significantly reduced in [X/H] vs FIP
 plot. Hence the possibility of outflowing magnetized column of gas
 systematically removing the singly ionized specie need to be explored.
 Another competing scenario is proposed by Luck \& Bond (1989) wherein the
 Lyman $\alpha$ photons from a shock in the atmosphere over ionized elements
 with SIP less than the Lyman limit, 13.6eV.
 While SIP plot displays a larger scatter than FIP; the later requires
 magnetic field driven wind  which are not easy to explain.
 In either case  a systematic time series spectral monitoring  to detect the events of
 strong shocks and  possible manifestation of stellar wind is required.

The long-term photometric monitoring from the All Sky Automated Survey (ASAS) shows
that IRAS13110\,-\,6629 and IRAS18489\,-\,0629 display periodic variations with characteristic
times between 52 and 60 days, with amplitudes in the 0.2-0.3 magnitude
range. The origin of these variations is most likely pulsational and are of a very
similar nature to those in other well known PAGB of intermediate temperature, such
as 89 Her, HD 161796 and LN Hya.
IRAS 17579\,-\,3121, IRAS 18321\,-\,1401 display fluctuations of a few hundredths of
magnitudes but no periodicities were found.
Multi-wavelength observations are required for better understanding of these objects.

\section*{Acknowledgments}

AAF acknowledges the support of DGAPA-UNAM grant through project IN104612. This work
has made extensive use of SIMBAD database, 2MASS (Two Micron All Sky Survey,
 which is a joint project of the University of Massachusetts and the Infrared
 Processing and Analysis Center/California Institute of Technology,
 funded by the National Aeronautics and Space Administration and the National Science Foundation),
 IRAS, AKARI/IRC data (a JAXA project
 with participation of ESA), WISE(Wide-field Infrared Survey Explorer,
 which is a joint project of the University of California, Los Angeles,
 and the Jet Propulsion Laboratory/California Institute of Technology,
 funded by the National Aeronautics and Space Administration),
 MSX6C( Midcourse Space Experiment  funded by the Ballistic Missile
 Defense Organization with additional support from NASA Office of
 Space Science)  and the ADS-NASA to which we are
 thankful. We would also like to express our gratitude to anonymous referee for
  his/her comments which have helped in improving the paper considerably.

\end{document}